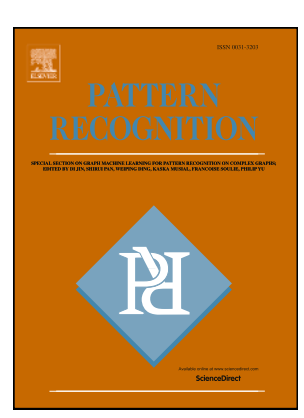

# Multi-view biclustering via non-negative matrix tri-factorisation

Ella S. C. Orme [a,*], Theodoulos Rodosthenous [a], Marina Evangelou [a]

[a] Department of Mathematics, Imperial College London, London, SW7 2AZ, United Kingdom

ABSTRACT

Multi-view data is ever more apparent as methods for production, collection and storage of data become more feasible both practically and fiscally. However, not all features are relevant to describe the patterns for all individuals. Multi-view biclustering aims to simultaneously cluster both rows and columns, discovering clusters of rows as well as their view-specific identifying features. A novel multi-view biclustering approach based on non-negative matrix factorisation is proposed named ResNMTF. Demonstrated through extensive experiments on both synthetic and real datasets, ResNMTF successfully identifies both overlapping and non-exhaustive biclusters, without pre-existing knowledge of the number of biclusters present, and is able to incorporate any combination of shared dimensions across views. Further, to address the lack of a suitable bicluster-specific intrinsic measure, the popular silhouette score is extended to the bisilhouette score. The bisilhouette score is demonstrated to align well with known extrinsic measures, and proves useful as a tool for hyperparameter tuning as well as visualisation.

## 1. Introduction

Biclustering, also known as two-way clustering, co-clustering and bi-dimensional clustering [1], refers to the simultaneous clustering of both the rows and columns of a data matrix. Initially popularised via application to gene expression matrices by Cheng and Church [2], it has seen success in many fields from climate science [3] to computer vision [4].

This manuscript proposes a multi-view biclustering approach, Restrictive Non-negative Matrix Tri-Factorisation, abbreviated as ResNMTF, based on non-negative matrix tri-factorisation (NMTF) [5]. NMTF is an extension of the well known non-negative matrix factorisation (NMF) method and has been applied as a single-view biclustering technique in many applications such as predicting drug synergism [6]. The intuitive interpretation of the factors that comes with their non-negativity is exploited to produce biclusterings. Further, the problem formulation easily allows for rows/columns to not belong to any bicluster (non-exhaustivity) and/or to belong to more than one bicluster (non-exclusivity/overlap).

Multi-view (also known as multi-modal) data refers to data collected from multiple sources describing the same objects, such as the same topics reported on by multiple news outlets or in multiple languages. Other examples include biomedical studies that commonly generate multiple large datasets describing the levels of gene expression (genomics), DNA methylation (methylomics) and proteins (proteomics), in a set of individuals or cells [7].

The benefits of working with multi-view data as opposed to single-view data, which include an ability to capture different aspects of the data, have been extensively discussed in the literature [8]. Rodosthenous et al. [9] have demonstrated that in comparison with analysing each view separately, integrating the different data sources can improve our understanding of the relationships found in the data, alongside increasing predictive accuracy. Additionally, using multi-view data can improve the stability of results [10] and provide more comprehensible visualisations than those obtained from single views alone [11].

When applied in a single, or multi-view setting, biclustering faces the same challenges as clustering including ensuring returned (bi)clusters are stable - meaning small perturbations in the input do not lead to large changes in the (bi)clustering. This problem is particularly pertinent when the number of true biclusters present is unknown. If too many biclusters are selected by an algorithm, or inputted by a user, these superfluous biclusters will be unstable. Analysis of stability of results provides a way to remove these falsely detected biclusters. Motivated by the work in von Luxburg [12], a stability analysis procedure is proposed that is implemented alongside ResNMTF. This stability analysis can be combined with other multi-view biclustering approaches for finding stable biclusters.

Another challenge inherited from clustering is the determination of the number of (bi)clusters. Whilst in clustering the quality of clusters is typically assessed using intrinsic measures, the structural complexity of biclusters means there is yet no standard internal biclustering measure [13]. An intrinsic measure should allow for non-exhaustivity and non-exclusivity. The application of biclustering methods in unsupervised settings, comparison of the methods themselves and comparison of results are therefore difficult.

* Corresponding author.
*E-mail address:* ella.orme18@imperial.ac.uk (E.S.C. Orme).

In this manuscript, an extension of the popular intrinsic measure silhouette score [14], named bisilhouette score, is proposed for biclustering. The bisilhouette score evaluates the obtained biclusters and can be used in the comparison between biclustering solutions, as well as for use in identification of the number of biclusters present. This latter application of the bisilhouette score is implemented in the proposed ResNMTF, with the optimal number of biclusters selected by maximising the bisilhouette score.

### 1.1. Overview/contributions

The manuscript is structured as follows. Section 2 presents multi-view biclustering approaches that exist in the literature and how they are related to the proposed ResNMTF. The novel intrinsic biclustering bisilhouette score is presented in Section 3. This is followed in Section 4 by the introduction of the multi-view biclustering method ResNMTF. An initialisation procedure based on singular value decomposition is proposed that is suitable for NMTF based methods. A stability analysis technique to improve the stability of returned biclusterings is introduced. The stability analysis can be combined with ResNMTF or other multi-view biclustering approaches for finding stable biclusters. The performance of ResNMTF is demonstrated, and compared with that of other methods, on real and synthetic datasets in Section 5. The efficacy of the bisilhouette score is also demonstrated on the real and synthetic data.

### 1.2. Notation

The following notation is used throughout the manuscript. Matrices are denoted by capital letters e.g. $X$ and are assumed to have rank greater than the number of biclusters considered. If $\mathcal{A}, \mathcal{B}$ are sets of natural numbers, $X_{.,\mathcal{A}}$ denotes the submatrix of X corresponding to all rows and the columns indexed by $\mathcal{A}$. Similarly, $X_{\mathcal{A},.}$ is the submatrix corresponding to the rows indexed by $\mathcal{A}$ and all columns. $X_{\mathcal{A},\mathcal{B}}$ is the submatrix corresponding to the rows in $\mathcal{A}$ and columns in $\mathcal{B}$. Unless otherwise specified $||\cdot||$ applied to a matrix denotes the Frobenious norm which is defined via $\|X\|^2 = \mathrm{tr}(X^T X) = \sum_{j=1}^{m} \sum_{i=1}^{n} (x_{ij})^2$. $||X_{.,k}||$ denotes the summation of the $k^{\text{th}}$ column of $X$ i.e. $||X_{.,k}|| = \sum_i X_{ik}$. The absolute function is denoted by $\mathrm{abs}(\cdot)$ and is applied elementwise to matrices. Set cardinality is represented by $|\cdot|$. For $a \in \mathbb{Z}_{\geq 1}$, $1:a$ denotes the set $\{1, \dots, a\}$. The functions $\mathrm{diag}(\cdot)$ and $\mathrm{bdiag}(\cdot)$ represent diagonal and block diagonal matrices respectively with the input along the diagonal. Data with $n_v$ views is represented by $\{X^{(v)}\}_{1:n_v}$ where $X^{(v)} \in \mathbb{R}^{n_r^{(v)} \times n_c^{(v)}}$. For view $v$ with $K$ biclusters, the biclustering is denoted by $\mathrm{BC}_K^{(v)} = \{(R_k^{(v)}, C_k^{(v)})_{1:K}\}$ where $R_k^{(v)}$ and $C_k^{(v)}$ are the sets of rows and columns belonging to bicluster $k$. Lower case $k$ generally denotes the $k^{th}$ bicluster whilst upper case $K$ refers to the number of biclusters present in a dataset. For a dataset with $n_v$ views, the biclustering is represented by $BC_K = \{\mathrm{BC}_K^{(v)}\}_{1:n_v}$.

## 2. Multi-view biclustering approaches

Various approaches have been considered to tackle the challenge of multi-view biclustering. These include methods based on; similarity scores [15], information theory [16], probabilistic models [17], graph theory [18] and deep learning [19]. In particular, methods based on dimensionality reduction techniques have proven popular [20–23]. This is partially due to the problem formulation which generally seeks an approximation of data as the product of two factors. Associating one factor with a row cluster and the other with a column cluster leads to a biclustering interpretation. By using shared factors across views, regularising factors towards each other, or using consensus factors, these methods can easily be extended to the multi-view setting.

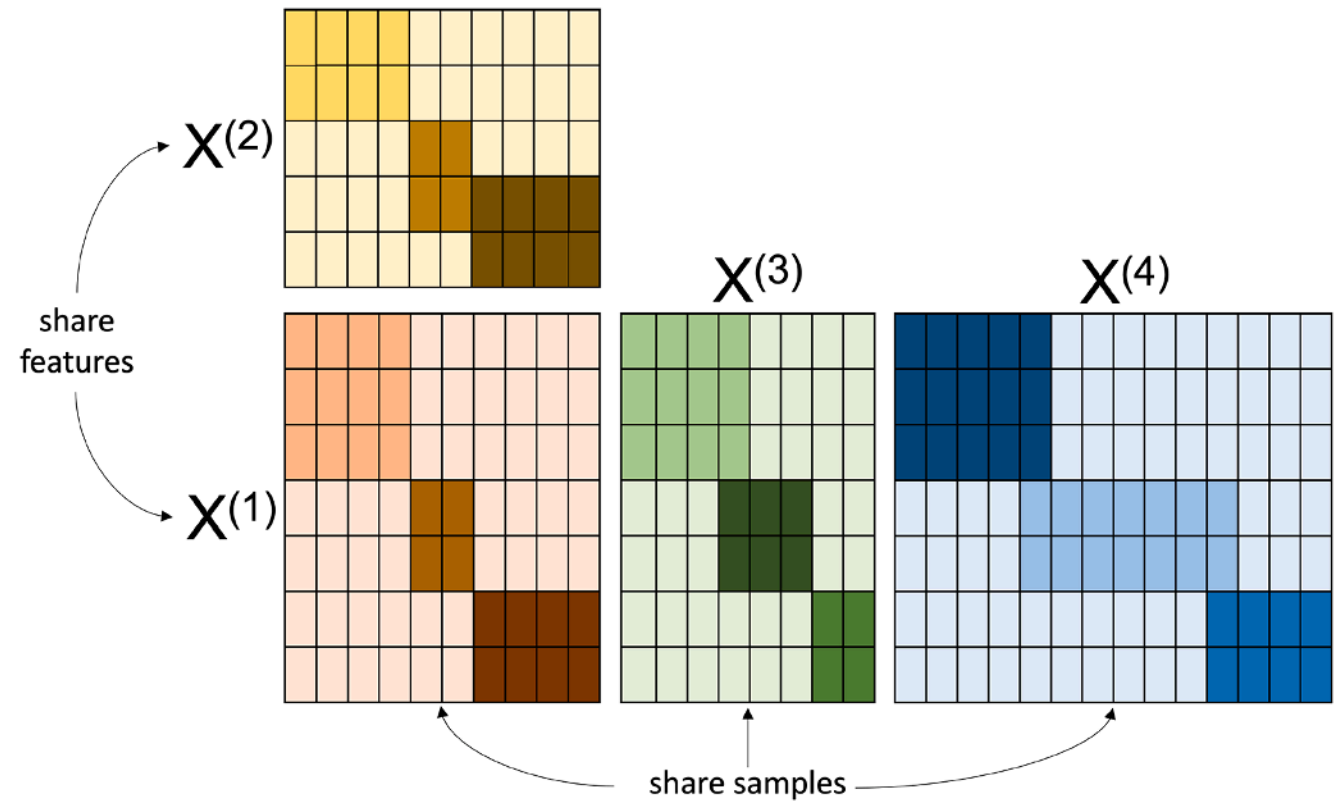

**Fig. 1.** Illustrative example of a scenario with a combination of shared rows and columns. $X^{(1)}$., $X^{(3)}$ and $X^{(4)}$ share the same individuals, representing e.g. three different omics datasets on the same individuals from one study. Features are shared by $X^{(1)}$ and $X^{(2)}$. This could represent the same features from a single omic (e.g. the same proteins) but with the two views obtained from two different clinical studies or on two different species. ResNMTF allows these relationships to be accounted for.

Most approaches seek a low rank matrix factorisation representation of the data (like the multi-view NMTF approach by Riddle-Workman et al. [24]), or are based on singular value decomposition, such as integrative sparse singular value decomposition (iSSVD) proposed by Zhang et al. [20]. iSSVD uses singular value decomposition to identify sparse rank one approximations to the data. By downweighting those samples already identified, biclusters are found sequentially without a reduced sample size or the need for pre-specifying the number of biclusters. This occurs alongside stability selection with the aim to control Type I error rates - aiding in the discovery of stable biclusters. Other methods, such as Group Factor Analysis (GFA) [21], take a generative Bayesian approach. By placing priors on the matrix factors and enforcing sparsity in both the samples and the features, GFA interprets the posteriors to obtain biclusters. In contrast to iSSVD, GFA starts with an overestimate of the number of biclusters and then removes those not deemed to be true biclusters. Both iSSVD and GFA allow for overlapping biclusters and non-exhaustivity. Both are intermediate multi-view biclustering approaches (*i.e.* they incorporate the multiple views within the models themselves) and are directly comparable approaches to the proposed ResNMTF.

A key limitation amongst the current dimensionality reduction based intermediate multi-view biclustering approaches is the lack of flexibility in the assumptions regarding relationships between views. Apart from GFA, other methods cannot enforce both shared rows and shared columns, nor a general combination of restrictions between views. There are various settings when this is desirable. Consider an example case with four views. If view 2 represents a repetition of an experiment represented in view 1, perhaps undertaken by a different research team but over the same genes, the columns associated to a bicluster in view 1 would be expected to be the same as those in view 2. Views 1, 3 and 4 could correspond to the same individuals (shared rows) but with each view representing a different omic. Fig. 1 illustrates this possibility of different view pairs sharing rows whilst other view pairs share columns. Some work, including iSSVD, enforces strict shared biclusters via shared consensus matrices [20,23]. This can lead to signal from individual views being ignored. It is not always advantageous to use all of the available views and careful selection of views is required in practice [11,25]. As this is difficult, enforcing strict shared biclusters leaves no room for suboptimal selection. Other methods require *a priori* knowledge of the number of clusters present [22], which is often not available.

ResNMTF addresses these limitations as it can be applied with any combination of restrictions between views, allowing shared row and column clusters to match prior knowledge, and without pre-existing knowledge of the number of biclusters. By regularising the factors towards each other across views ResNMTF avoids issues with noisy views skewing results caused by the use of consensus or shared matrices.

## 3. Bisilhouette score

Assessing the obtained biclusters and finding the optimal number of biclusters is an inherited problem from clustering. The additional structural complexity of biclusters exacerbates this issue. Existing literature lacks a bicluster specific approach that considers the factors usually deemed necessary for assessing clusters: compactness and degree of separation. Such measures in addition need to allow for non-exhaustivity and non-exclusivity of biclusters.

Current bicluster literature focuses on extrinsic measures that use external information to assess the quality of biclusters [26] or coherence measures [27] that check how closely a bicluster adheres to a specific structure (e.g. constant, coherent). These latter intrinsic measures generally consider biclusters in isolation and ignore any relationships between them. The exception is the Relevance Index (RI) proposed by Yip et al. [28] which considers differences in local and global variation along the columns belonging to a bicluster. However this measure does not penalise for the presence of redundant columns, meaning the presence of irrelevant columns in a bicluster is not necessarily penalised.

Biclusters are often evaluated as two separate sets of clusters (row and column clusters) via standard clustering measures, a process that risks two issues. Firstly, promising scores can be returned when the rows of a cluster are successfully identified but associated columns are incorrectly or poorly identified (including incorrect matching between row and column clusters if method appropriate). Secondly, a poor score may be returned to a correctly identified bicluster if the bicluster columns are a small portion of the overall matrix. A marked difference across the rows over these columns may be obscured when all columns are considered. The literature lacks a bicluster specific intrinsic measure, a gap that we fill with our bisilhouette score proposal. The proposed measure has the following desirable properties: works for general unsupervised settings, works in the presence of overlapping and/or non-exhaustive biclusters, as well as incorporating both the structure of the bicluster and how it relates to the rest of the data.

The proposed bisilhouette score is based on the popular intrinsic clustering measure, the silhouette score [14], that takes into account compactness and separation of clusters as well as allowing for overlap and non-membership. The silhouette score considers the average distance (typically Euclidean) between an element and others belonging to the same cluster as well as the average distance to the nearest cluster. More specifically, for element $i$ in cluster $k$ the average distance to the other elements in the same cluster is given by $a_i = (|R_k| - 1)^{-1} \sum_{l \in R_k} d_{il}$, where $d_{jl}$ is the Euclidean distance between row $j$ and row $l$ of the data matrix. The average distance to the elements in the next closest cluster is calculated via $b_i = \min_{j \in 1:K, j \neq k} |R_j|^{-1} \sum_{l \in R_j} d_{il}$. The silhouette coefficient $s_i$ for element $i$ is defined by $s_i = (b_i - a_i)/\max(b_i, a_i)$.

The original silhouette score takes the average of the $s_i$ over all datapoints as an overall score of the clustering. However this does not allow for overlapping clusters. Therefore to calculate the final score, we obtain scores for each cluster by averaging over the silhouette coefficients corresponding to the elements in that cluster. These are further averaged to give an overall score for the clustering as in the following equation:

$$S = \frac{1}{K} \sum_{k=1}^{K} \frac{1}{n_k} \sum_{i \in R_k} s_i \tag{1}$$

### 3.1. Bicluster extension

To extend the silhouette score to the bicluster case, the silhouette coefficients are calculated for each row in a given bicluster, based only on the columns corresponding to that bicluster. The score for bicluster $k$ ($B_k$) is obtained as follows: the data matrix $X$ is subsetted by the columns belonging to $C_k$, $\tilde{X} = X_{\cdot,C_k}$. Treating $R_j$ as the clusters, the silhouette coefficients for the elements of $R_k$ on $\tilde{X}$ are calculated. Averaging over the coefficients, $B_k$ is found as illustrated in Fig. 2.

As with the silhouette score, $B_k$ takes values in $[-1, 1]$ with a higher score indicating a more compact and well separated bicluster. A value of 1 is obtained if the columns are constant within the bicluster. Care should be taken as this can be achieved when the columns are constant across all rows, not just those belonging to the bicluster. This issue is prevented by removing features with zero/very low variance during the pre-processing stage. An overall bisilhouette score $B$ for the biclustering $\{(R_k, C_k)\}_{1:K}$ is given by calculating the mean over the non-zero $B_k$. A score of zero is assumed to correspond to an empty bicluster - a non-empty bicluster with score exactly zero is very unlikely for data with any signal present. The above assumes all $R_k$ and $C_k$ are non-empty and there are at least three unique sets amongst the row clusters. The case where there are fewer than three unique row clusters is treated separately in Supplementary Material (S) 1.1. With these edge cases considered, the bisilhouette score can be calculated for any number of biclusters, allowing for comparison between any two sets of biclustering results. This is in contrast to the silhouette score which is only defined for two or more clusters.

## 4. Restrictive non-negative matrix tri-factorisation

This section introduces the novel biclustering approach, ResNMTF, that is based on non-negative matrix tri-factorisation. The aim of the proposed approach is to identify biclusters amongst the integrated dataviews whilst allowing for any non-exhaustivity and/or non-exclusivity to be present. The approach is combined with a stability analysis technique that checks the stability of the selected biclusters.

### 4.1. Non-negative tri-matrix factorisation and extensions

The standard NMTF approach factorises the non-negative design matrix $X \in \mathbb{R}^{N \times p}$ into a product of three matrices (as opposed to two factors in NMF), such that $X \approx FSG^T$ where $F \in \mathbb{R}^{N \times K}$, $S \in \mathbb{R}^{K \times L}$ and $G \in \mathbb{R}^{p \times L}$. Here $K$ denotes the number of row clusters and $L$ denotes the number of column clusters, with $K = L$ often chosen.

Multi-view NMTF approaches seek matrix factorisations for each view $v \in 1 : n_v$, $X^{(v)} \approx F^{(v)} S^{(v)} (G^{(v)})^T$, with appropriate dimensions implied for all matrices. Like in multi-view NMF methods (which seek factorisations $X^{(v)} \approx F^{(v)} (W^{(v)})^T$), commonality across the views can be implemented via different means. Akata et al. [29] implement a shared $F$ matrix, setting $F = F^{(v)}$ for all $v$. Others regularise the matrices corresponding to row cluster assignments for each view ($F^{(v)}$) towards a single consensus matrix $F^*$ [30]. CoNMF proposed by He et al. [31] regularises factors towards each other via penalty terms such as $\|F^{(v)} - F^{(w)}\|$. By clustering with individual factors for each view CoNMF aims to alleviate the impact on performance a single poor quality view has [31]. Yang et al. [32] proposed CoNMTF that extends CoNMF to the biclustering case where similarly to other NMTF multi-view methods [24,33], it is based on relational matrices and is therefore not applicable in a general setting.

ResNMTF, similarly to CoNMTF, extends CoNMF to the biclustering case but in contrast to coNMTF is suitable for application with general multi-view data. The regularisation of factors implemented in ResNMTF tackles a key difficulty associated with multi-view integration methods. As it is not possible to know which are the noisy views and which are the informative ones, a desirable property of a proposed method is its ability to prevent noisy views from corrupting the signal of the informative ones. By regularising the factors towards each other across views, the relationships between views are accounted for, whilst avoiding issues with noisy views skewing results caused by the use of consensus or shared matrices [30].

A multi-view NMTF approach to multi-view biclustering inherits several problems from both NMF and NMTF. Firstly, as with NMF, re-strictions on matrix factors are needed to address problems of non-identifiability. Additional constraints are required moving from NMF to NMTF to prevent reformulation as an NMF solution, for example with the two factors $F$ and $SG$. Such constraints include orthogonal constraints on both $F$ and $G$ [5]. As with all methods involving update-based optimisation procedures, like NMF, initialisation procedures are required - this motivates our proposal of an initialisation procedure suitable for NMTF based methods in Section 4.4. Lastly, NMTF requires pre-specification of the number of biclusters ($K$). When prior information is available regarding the true number of biclusters, this value of $K$ can be used to obtain the result. However, in most unsupervised learning cases the optimal value of $K$, $\hat{K}$, is unknown. Within ResNMTF various possible values of $K$ are considered and the bisilhouette score determines the optimal value. This pre-specification of $K$ does not allow for all possible scenarios. Suppose a particular view doesn't contain any biclusters, simple NMTF based approaches will still return biclusters for this view. In light of this, ResNMTF (via a resampling approach) assesses each bicluster and removes those that are deemed to represent noise rather than signal. This removal, alongside the stability analysis procedure, ensures only stable biclusters representing true signal are returned by ResNMTF.

**Fig. 2.** Schematic for the calculation of the bisilhouette score. Matrix $X$ is subsetted by considering the columns belonging to the first bicluster ($C_1$), forming $\tilde{X}$.. The silhouette score on $\tilde{X}$ is calculated for the rows belonging to the first bicluster ($R_1$). These values are averaged to give a score for ($R_1, C_1$).

The section starts with a presentation of the optimisation problem with the set of update steps taken as well as the proposal of the initialisation technique. A discussion of how biclusters are assigned amongst the integrated data-views is presented. Next, the steps taken for removing spurious biclusters and checking the stability of the selected ones are presented. The section ends with the implementation of the approach.

### 4.2. Objective

By assuming a known $K$, $K \geq 3$, the aim of Restrictive Non-negative Matrix Tri-Factorisation is to solve the optimisation problem given by:

$$\min_{F^{(v)},S^{(v)},G^{(v)}} \left( \sum_{v=1}^{n_v} \left\| X^{(v)} - F^{(v)}S^{(v)}\left(G^{(v)}\right)^T \right\|^2 + \sum_{w=v+1}^{n_v} \sum_{v=1}^{n_v-1} \left( \phi_{vw} \left\| F^{(v)} - F^{(w)} \right\|^2 + \xi_{vw} \left\| S^{(v)} - S^{(w)} \right\|^2 + \psi_{vw} \left\| G^{(v)} - G^{(w)} \right\|^2 \right) \right) \tag{2}$$

such that for all $v$ in $1 : n_v$

- $||F^{(v)}_{\cdot,k}||, ||G^{(v)}_{\cdot,k}|| = 1, \quad \forall k = 1, 2, \dots, K$
- $F^{(v)}, S^{(v)}, G^{(v)} \geq 0$

where $\Phi = (\phi)_{vw}$, $\Xi = (\xi)_{vw}$, $\Psi = (\psi)_{vw}$ are upper triangular non-negative matrices. $\phi_{vw}, \xi_{vw}, \psi_{vw}$ are tuning parameters allowing for different restrictions. Different combinations of tuning parameters enforce different combinations of restrictions. For example, non-zero $\phi_{12}$ means views one and two share common row clusters, therefore $F^{(1)}$ and $F^{(2)}$ are forced towards each other. After minimising the objective function defined in Eq. (2), factors $F^{(v)}, S^{(v)}$ and $G^{(v)}$ for each view $X^{(v)}$ are obtained. The factors for view $v$ determine biclusters $BC^{(v)}_K$, giving the multi-view biclustering result, $BC_K$.

Including the regularisation of both $S^{(v)}$ and $G^{(v)}$ matrices in the objective function allows for different biclustering scenarios. Such scenarios include common column clustering across views rather than, or in addition to, across views.

### 4.3. Update steps

The objective function of NMTF is non-convex and minimising it has been shown to be an NP-hard problem [34]. However, by fixing all matrices except one, the problem becomes convex in the given matrix. Following the work by Long et al. [35] and He et al. [31], which are themselves based on the algorithm by Lee and Seung [36], a local solution to this problem is found via multiplicative update rules. The update rules move the variables along the gradient direction. This involves initialising factors $F^{(v)}, S^{(v)}$ and $G^{(v)}$ for each view and iteratively applying multiplicative updates to each matrix, until some predetermined notion of convergence is met. As our objective function requires normalisation and non-negativity constraints to be met, Lagrangian multipliers $\lambda^{(v)}_k$ and $\mu^{(v)}_k$ are added to the objective function that are also iteratively updated. The update steps are as follows:

$$F^{(v)} \leftarrow F^{(v)} \odot \frac{X^{(v)}G^{(v)}\left(S^{(v)}\right)^T + \sum_{u=1}^{n_v} \phi_{uv} F^{(u)}}{F^{(v)}S^{(v)}\left(G^{(v)}\right)^T G^{(v)}\left(S^{(v)}\right)^T + \left(\sum_{u=1}^{n_v} \phi_{uv}\right) F^{(v)} + \frac{1}{2}\Lambda^{(v)}}$$

$$S^{(v)} \leftarrow S^{(v)} \odot \frac{(F^{(v)})^T X^{(v)} G^{(v)} + \sum_{u=1}^{n_v} \xi_{uv} S^{(u)}}{\left(F^{(v)}\right)^T F^{(v)}S^{(v)}\left(G^{(v)}\right)^T G^{(v)} + \left(\sum_{u=1}^{n_v} \xi_{uv}\right) S^{(v)}}$$

$$G^{(v)} \leftarrow G^{(v)} \odot \frac{(X^{(v)})^T F^{(v)} S^{(v)} + \sum_{u=1}^{n_v} \psi_{uv} G^{(u)}}{G^{(v)}\left(S^{(v)}\right)^T \left(F^{(v)}\right)^T F^{(v)}S^{(v)} + \left(\sum_{u=1}^{n_v} \psi_{uv}\right) G^{(v)} + \frac{1}{2}\mathbf{1}M^{(v)}}$$

$$\Lambda^{(v)} \leftarrow \Lambda^{(v)} \odot (\mathbf{1}^T F^{(v)})$$

$$M^{(v)} \leftarrow M^{(v)} \odot (\mathbf{1}^T G^{(v)})$$

where $\Lambda^{(v)} = (\lambda^{(v)}_1, \dots, \lambda^{(v)}_K)$ and $M^{(v)} = (\mu^{(v)}_1, \dots, \mu^{(v)}_K)$. Here $\mathbf{1}$ is a column vector of appropriate length for multiplication. A full derivation of the update steps can be found in Section S1.2.

For iteration $i$, the error is given by:

$$E_i = \frac{1}{n_v} \sum_{v=1}^{n_v} E_v \text{ where } E_v = \|X^{(v)} - F^{(v)}S^{(v)}\left(G^{(v)}\right)^T\|^2 / \|X^{(v)}\|^2 \tag{3}$$

A convergence based stopping criteria with a tolerance of $e^{-6}$ between successive errors is implemented as suggested in Čopar et al. [37].

#### 4.3.1. Computational complexity

The computational complexity of the update steps is discussed using big $O$ notation. The complexity for each matrix factor is broken down into the complexity of the numerator and denominator in the update step. For simplicity, here and throughout Sections 4.4 and 4.5, we drop the view subscript index. We assume $K, n_v \ll n_r^{(v)}, n_c^{(v)}$ for all $v$. Let $N$ and $p$ be the maximum number of rows and columns respectively across views.

For matrix $S$, the numerator has complexity, $O(NpK) + O(NK^2) + O(n_v K^2) \approx O(NpK)$ whilst the denominator is $O(pK^2) + O(NK^2) + O(2K^3) \approx O((N+p)K^2)$. The element-wise multiplication and division adds $O(2K^2)$ operations, giving an overall complexity of $O(NpK) + O((N+p)K^2) + O(2K^2) \approx O(NpK)$. Considering matrix $F$, the numerator has complexity $O(NpK) + O(NK^2) + O(n_v NK) \approx O(NpK)$ whilst the denominator is $O(2NK^2) + O(pK^2) + O(K^3) + O(NK) + O(1) \approx O((2N+p)K^2)$. Multiplication and division introduce an extra $O(2NK)$ operations resulting in an overall complexity of $O(NpK) + O((2N+p)K^2) + O(2NK) \approx O(NpK)$. Similar computations give a complexity of $O(NpK)$ for matrix $G$. As the Lagrange multipliers updates have $O(n_v(N+p)K)$ complexity, the cost of one iteration of ResNMTF is $O(n_v NpK)$. This is the same as applying NMF to each view independently and to the popular multi-view NMF approach, MultiNMF [30].

### 4.4. Initialisation

As non-convexity implies local solutions to NMTF, the initialisation of factors is extremely important in the success of the method. The proposed initialisation strategy builds upon the work of Boutsidis and Gallopoulos [38] and Riddle-Workman et al. [24] for NMTF based methods and is applied to each view separately. The proposed initialisation is suitable for general matrices and provides a different initialisation for each view in contrast to the work of Riddle-Workman et al. [24].

Let $\sigma_k$ be the non-negative values that correspond to the singular values of $X$ in decreasing order and $\boldsymbol{u}_k$ and $\boldsymbol{v}_k$ be the corresponding left and right singular vectors respectively. The initial values of the multipliers $\Lambda$ and $M$ are $\Lambda = M = \mathbf{1}^T$.

$F$ and $G$ are initialised as:

$$F \leftarrow (\text{abs}(\boldsymbol{u}_1), \ldots, \text{abs}(\boldsymbol{u}_K))\text{diag}\left(\{1/||\boldsymbol{u}_k||\}_{1:K}\right)$$

$$G \leftarrow (\text{abs}(\boldsymbol{v}_1), \ldots, \text{abs}(\boldsymbol{v}_K))\text{diag}\left(\{1/||\boldsymbol{v}_k||\}_{1:K}\right)$$

Both matrices are normalised so that they satisfy the necessary normalisation constraints presented earlier.

$S$ is initialised as the sum of the diagonal matrix of the first $K$ singular values plus a $K \times K$ matrix of noise with entries from $N(0, \sigma_N^2)$. The multiplicative nature of the updates require the addition of positive noise to allow for non-zero off-diagonal entries in $S$ and $\sigma_N$ denotes our belief that the biclusters are distinct. $S$ then absorbs the normalisation factors from $F$ and $G$ giving:

$$S \leftarrow \text{diag}\left(\{||\boldsymbol{u}_k)||\}_{1:K}\right)(\text{diag}(\{\sigma_k\}_{1:K}) + \text{abs}(N))\text{diag}\left(\{||\boldsymbol{v}_k||\}_{1:K}\right) \quad (4)$$

### 4.5. Bicluster assignment

The factors obtained from the optimisation outlined in Eq. (2) are used to define the biclusters. Let $\tilde{R}_k$ be the row cluster associated with the $k^{\text{th}}$ column of $F$. As $F_{ik}$ represents the association of row $i$ to $\tilde{R}_k$ and column normalisation requires $||F_{.k}|| = 1$, the following relation determines row clusters assignment:

$$\tilde{R}_k = \left\{ i : F_{ik} > \frac{1}{n_r} \right\}$$

If $X$ is pure noise, all rows would be equally weakly associated to a given row cluster. Therefore row $i$ is assigned to $\tilde{R}_k$ if it is more strongly associated to the cluster than expected if the data was purely noise.

Similarly, $G_{ik}$ represents the association of column $j$ to column cluster $k$ ($C_k$) and so column clustering is determined via:

$$\tilde{C}_k = \left\{ i : G_{ik} > \frac{1}{n_c} \right\}$$

As all row and column clusters are treated independently, this assignment technique allows for non-exhaustive and non-exclusive biclusterings. Additionally, this cluster assignment allows for complete overlap in either dimension. Finally, $BC_k$ is defined by matching $C_k$ to the row cluster $\tilde{R}_l$ via $l = \text{argmax}_j\, S_{jk}$, resulting in $R_k = \tilde{R}_l$.

### 4.6. Removal of spurious biclusters

Following the assignment of biclusters, the results are checked to ensure the biclusters are a true signal. Spurious biclusters are identified and removed from the biclustering. Biclusters are classified as spurious ones if they do not differ from what is expected from noise. The following re-sampling procedure is applied for identifying the spurious biclusters. For resampling $m$, the entries of the original data matrix $X^{(v)}{}_{1:n_v}$ are shuffled creating a pure noise data matrix, $\hat{X}^{(v)}_{m,1:n_v}$. The objective function in Eq. (2) is solved for the shuffled data-views and the factors $\hat{F}^{(v)}_{m,1:n_v}$ are obtained. For each $v$, $\hat{X}^{(v)}_m$ is pure noise and all rows are equally weakly associated to each bicluster. Therefore each entry of $\hat{F}^{(v)}_m$ is independent and identically distributed. These column vectors will be differently distributed to those from the original factors corresponding to true biclusters (as illustrated in Figure S2). The Jensen Shannon divergence (JSD) [39] (see Section S1.3.1 for details) is used to test the similarity between column vectors obtained from the true data vector and the noisy one. When the distribution of the column vector does not differ significantly from that obtained from the noisy data, the bicluster is deemed false and removed. For every $v$, each column of $F^{(v)}$ is considered independently, allowing for any particular bicluster to be removed. Consider the multi-omics example where biclusters correspond to disease subtypes; a particular subtype may have distinguishing features in a certain omic but not in others. Examining columns of $F^{(v)}$ separately allows for the spurious biclusters found in the irrelevant omics to be removed. The method is specified in detail in Section S1.3 with pseudocode provided in Algorithm S1.

### 4.7. Determining the number of biclusters

Biclustering is often applied in unsupervised settings and hence the number of biclusters present is unknown. In order to determine the optimal number of biclusters $\hat{K}$, ResNMTF considers a range of possible values $K \in (K_{\min}, K_{end})$. The procedure for removal of spurious biclusters is performed for each value of $K$, returning a biclustering $BC_K$. The bisilhouette score for each $BC_K$ is calculated ($B_K$) and used to select between biclusterings. When the bisilhouette score is maximised at either limit of $k$ ($\hat{K} = K_{\min}$ or $K_{end}$), the current implementation of ResNMTF extends the range of values considered until this is not the case.

### 4.8. Stability analysis

von Luxburg [12] define the *instability of a clustering algorithm* as the expected distance between two clusterings produced from two different samples of the same size. The same principle applies in biclustering. Even after removing spurious biclusters, unstable biclusters may remain. Motivated by the general algorithm for calculating a stability score in von Luxburg [12] as well as the work in Duong-Trung et al. [40], a strategy is implemented for assessing the biclusters through a sub-sampling technique. After applying the sub-sampling method introduced below, the results are compared to the results of the original data, and any biclusters with consistently low similarity to the resampled results are deemed unstable and removed.

The re-sampling procedure is described here. The $m^{th}$ resampled dataset is produced by randomly sub-setting matrix $X^{(v)}$, producing submatrix $\tilde{X}^{(v)}_m$ consisting of $\lfloor \alpha n_r^{(v)} \rfloor$ rows and $\lfloor \alpha n_c^{(v)} \rfloor$ columns. Here $\alpha$ is the sample rate (default value 0.9) and denotes the portion of the original data in the resampled dataset. If shared row and/or column cluster restrictions are enforced, this is imposed here with the resampling consistent across the relevant dimension and views. The same subsampling is applied to the original biclustering producing $\mathrm{BC}_m^{(v)}$, allowing for meaningful comparisons. ResNMTF is performed on $\{\tilde{X}_m^{(v)}\}_{1:n_v}$ with the number of biclusters fixed and equal to $\hat{K}$, producing biclustering $\tilde{\mathrm{BC}}_m$. Since the 'ground truth' labelling $(\mathrm{BC}_m^{(v)})$ is available , the extrinsic measure relevance (defined in Section S2.1) is used to compare the results achieved from the subsample with those from the original data to produce a similarity score for each bicluster $l$ in each view, $\mathrm{Rel}_{lm}^{(v)}\left(\mathrm{BC}_m^{(v)}, \tilde{\mathrm{BC}}_m^{(v)}\right)$. This is repeated $n_s$ times and the mean score across repetitions $(\bar{\mathrm{Rel}}_l^{(v)})$ is calculated and if $\bar{\mathrm{Rel}}_l^{(v)} \leq \omega$ for some threshold $\omega$, bicluster $l$ in view $v$ is deemed unstable and the update $R_l^{(v)} = C_l^{(v)} = \emptyset$ is performed. The number of repetitions controls the number of subsampled datasets used to estimate the average, with $n_s = 5$ used as default. The threshold $\omega$ controls the balance between recovering all biclusters to some degree and ensuring those biclusters returned are stable. It is a hyperparameter to be tuned and the choice depends on the application setting and the desired properties of the results.

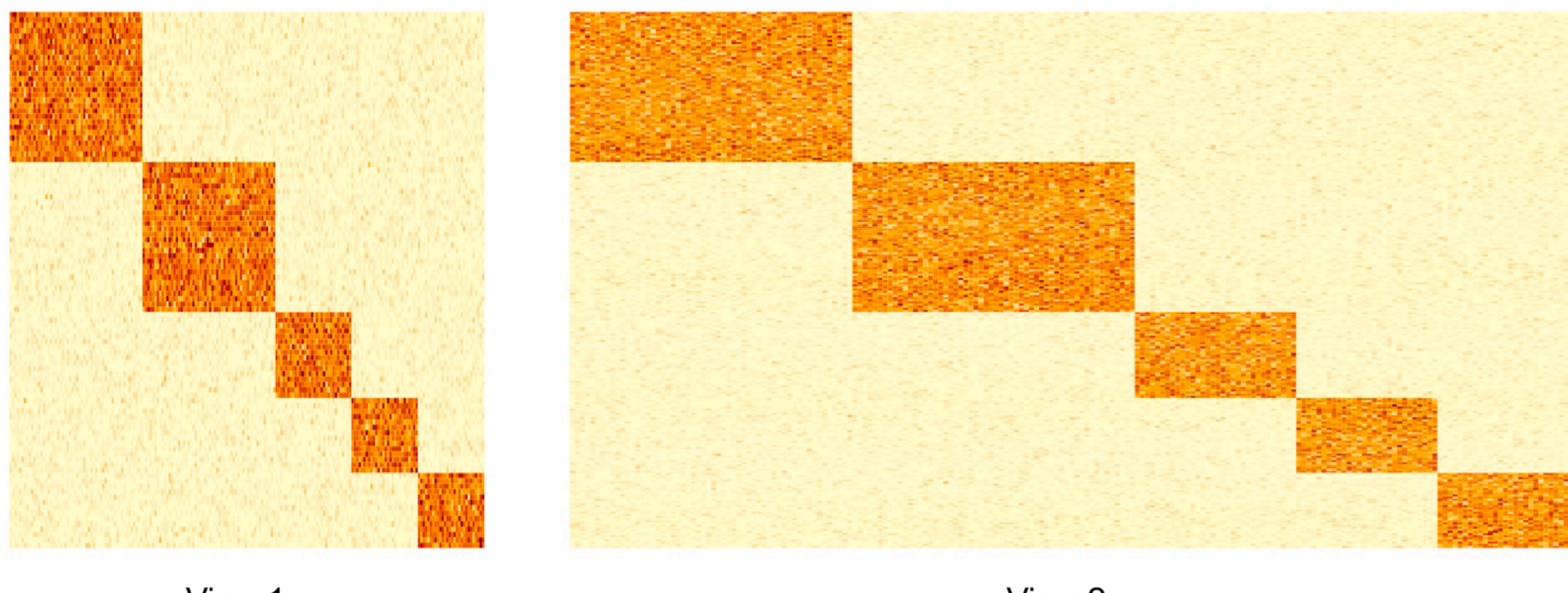

**Fig. 3.** Synthetic data example. Column cluster dimensions are $(27, 27, 15, 13, 13)$ in view 1 and $(69, 69, 39, 34, 34)$ in view 2.

### 4.9. Implementation

Pseudocode describing the implementation of stability analysis within ResNMTF is provided in Algorithm S2 and the application of ResNMTF is summarised in Algorithm S3. The R packages 'resnmtf' and 'bisilhouette', implementing ResNMTF and the bisilhouette score respectively, are available at https://github.com/eso28599/resnmtf and https://github.com/eso28599/bisilhouette. The code required to reproduce the results from the paper can be found at https://github.com/eso28599/resNMTF_paper.

## 5. Numerical experiments

This section demonstrates the performance of ResNMTF and the efficacy of the bisilhouette score as an intrinsic measure on both synthetic and real datasets with different characteristics. ResNMTF is compared with the unrestricted NMTF, implemented via $\Phi = \Psi = \Xi = 0$. Unrestricted NMTF is applied in a similar fashion to ResNMTF, following the steps described in Section 4. Further, ResNMTF is compared with the competing methods iSSVD and GFA which were introduced in Section 2 (see Section S2.5 for implementation details).

The performance of the multi-view biclustering approaches is assessed by calculating both extrinsic and intrinsic measures. The Jaccard [41] based measures relevance and F-score (defined for biclusters in Section S2.1) are used to evaluate the quality of returned biclusters, comparing the reported biclusters with the underlying true biclusters. Our proposed intrinsic measure, the bisilhouette score (abbreviated as BiS) is also reported. Throughout the conducted experiments, these scores are calculated for each view individually and averaged. For both measures a score closer to 1 indicates better performance.

The section is structured as follows. A description of the synthetic and real data analysed is provided. Subsequently a discussion on how the hyper-parameters, $\Phi, \Psi$ and $\omega$, of ResNMTF are tuned is presented. The section ends with a presentation of the findings from the conducted analyses. The conducted analyses are used to address the following key questions:

1. Is the bisilhouette score a powerful measure to identify the correct number of biclusters?
2. How does the bisilhouette score compare with other intrinsic and extrinsic measures for evaluating biclusters?
3. Is ResNMTF able to correctly identify the biclusters of the data?
4. How does ResNMTF compare with other biclustering approaches?

### 5.1. Data

Synthetic data were created where the true structure is known to assess the performance of ResNMTF under different scenarios including number of views, biclusters or individuals, as well as noise levels.

For each view $v$, the data matrix $X^{(v)}$ is generated as the sum of the absolute values of a signal matrix, $B^{(v)}$, and a noise matrix, $\mathcal{E}^{(v)}$. The absolute values are used to ensure non-negativity of the data. The entries of the noise matrix $\mathcal{E}^{(v)}$ are drawn from a $N(0, \sigma^2)$ distribution, with the $\sigma$ parameter controlling the level of noise in the data. In the conducted experiments the default value of $\sigma^2$ was 5, and in the experiments where the effect of noise was investigated the value of $\sigma^2$ was varied between 1 to 100.

Each bicluster is represented by a block matrix, $B_k^{(v)}$, along the diagonal of $B^{(v)}$ i.e. $B^{(v)} = \mathrm{bdiag}(\{B_k^{(v)}\}_{1:K})$, as illustrated in Fig. 3. $B_k^{(v)}$ has $r_k$ rows and $c_k$ columns. The entries of these blocks are samples from a Normal distribution with mean $\mu$ and variance $\sigma_B^2$ $(N(\mu, \sigma_B^2))$, where in the conducted experiments the parameters were set up as 5 and 1 respectively. As $\sigma_B$ is independent of $k$ and $v$, the within-bicluster variation is constant both across a view and between all views. The exhaustive exclusive biclusters are generated in such a way so that the row clusters are shared across the views but separate column clusters exist. For a given number of rows/columns and number of biclusters, the dimensions $((r_1, \ldots, r_K)/(c_1, \ldots, c_K))$ were fixed as follows for consistency. Biclusters are ordered in decreasing size, with an initial ratio of 4:3 for $K = 2$. As the number of biclusters increases by one, the part corresponding to the current largest bicluster splits in two, with $2r + 1$ being split into $r + 1$ and $r$. Any remainders from scaling the ratios are added to the first of the smallest biclusters. For example, if $K = 5$, $n_v = 2$, and $n_r = 200$, the row clusters are $(55, 55, 31, 27, 27)$ (as illustrated in Fig. 3). The number of individuals is fixed in most experiments at $n = 200$, except when this is the factor being investigated in which case it varies from 50 to 1000. Similarly, most experiments are conducted on three views with 100, 50 and 250 features in each view, with the exception of the study on increasing number of views when all views have 150 features. As a final step, the rows and columns of $X^{(v)}$ are shuffled to more accurately represent expected data. The same row shuffling is applied across all views whilst column shuffling is done independently. Amendments to this process to produce overlapping and/or non-exhaustive biclusters are outlined in Section S2.2. Results from additional experiments investigating; the effect of the rate of overlap, non-exhaustivity and the studies outlined here for two alternative base scenarios, can be found in Section S2 with specific results pointed to throughout this section. Further, we have generated synthetic data using the approach outlined by [20], the details and results of which are presented in Section S2.6.2.

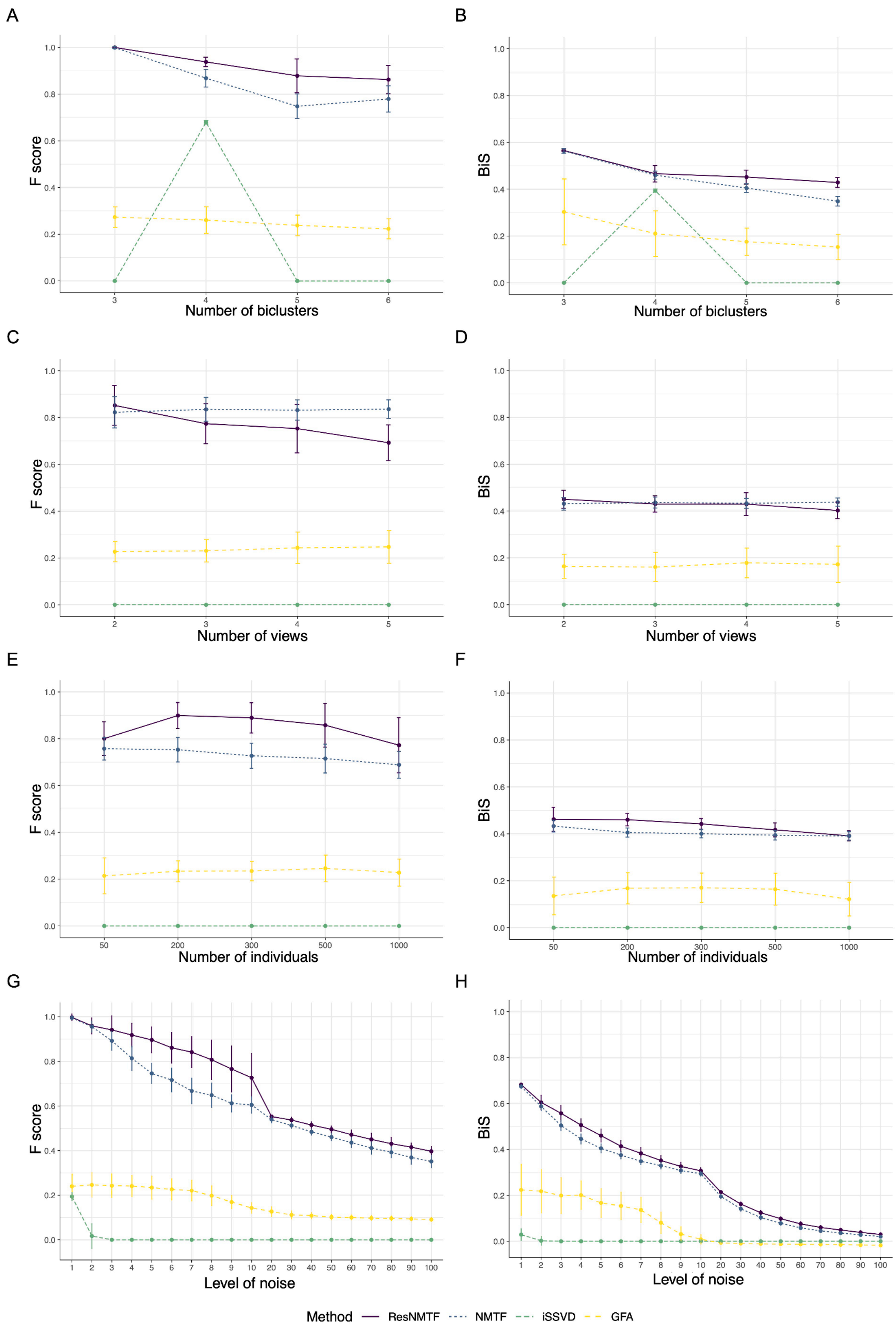

**Fig. 4.** F-score (left column) and BiS (right column) for different methods: base scenario of 200 rows, 3 views (100, 50 and 250 columns), 5 biclusters and $\sigma = 5$. The number of (A-B) biclusters (C-D) views (150 columns each) (E-F) individuals or (G-H) the level of noise are varied. The average score across 100 repetitions is reported with $\pm$ standard deviation error bars presented.

**Table 1**
Summary of the real datasets used including the number of true row clusters present, the number of views in each dataset, the number of samples (which is equal for each view) and the number of features in each view.

| Datasets | Clusters | View | Samples | Features |
|---|---|---|---|---|
| 3Sources | 6 | 3 | 169 | (3393, 3553, 2998) |
| BBCSport | 5 | 2 | 544 | (3173, 3195) |
| A549 | 3 | 2 | 2641 | (1176, 1147) |
| TCGA | 3 | 2 | 1241 | (20267, 631) |

The following real multi-view datasets, summarised in Table 1, are also analysed. The pre-processing steps applied are outlined in Section S2.3.

- **3Sources**[1] is a popular dataset that includes articles on the same topics reported by three different news outlets (corresponding to the three separate views); the Guardian, BBC, and Reuters. Each view is a document-term matrix presenting 169 articles (documents) and the counts of various words (terms) across the articles. There are 3393, 3553 and 2998 terms in each of the views respectively. Each article was manually labelled by the source as one or more of six categories; business, entertainment, health, politics, sport or tech.
- **BBCSport**[2] contains 169 articles which have been split into two parts (corresponding to two views). Each article has been labelled in one of the five categories: athletics, cricket, football, rugby, tennis. The two views contain 3173 and 3195 terms, respectively.
- **A459**[3] is a single cell multi-omic dataset obtained from a kidney tissue of a mouse [42]. The subtypes represent differential treatment time of the cells of either 0, 1 or 3 h. Transcriptomic (scRNA-seq) and epigenomic (scATAC-seq) data are obtained on the 2641 cells, with the views containing 1176 and 1147 genes respectively.
- **TCGA**[4] is a multi-view dataset that has been constructed from the TCGA study [7] and consists of gene expression and miRNA data from individuals with one of three types of cancer (AML, breast or colon). Only the genes analysed in all three cancer studies, and with non-zero value in at least one individual, are considered. This resulted in 1241 individuals with 20,267 and 631 genes in views one and two respectively.

Most of the datasets are sparse with 3Sources and BBCSport containing non-zeros in less than 5 % of entries, and for A549, in less than 25 %. Euclidean distance is not typically used for sparse data with the Manhattan or cosine distances preferred. Whilst for A549 the performance with different distances is comparable, for the two datasets with over 95 % sparsity an improvement is observed with the cosine distance (Table S1). The use of cosine distance also improves performance of ResNMTF on the TCGA dataset (Table S1). In the subsequent analyses the bisilhouette score is therefore implemented with the cosine distance for the 3sources, BBCSport and TCGA datasets.

**Table 2**
F-score of ResNMTF and competing methods on real datasets. Optimisation measure of restriction and stability hyperparameters denoted in brackets e.g. (BiS/F) means $\phi/\psi$ is selected by maximising the bisilhouette score and $\omega$ is selected by maximising F-score. Best results in bold.

| Method | 3Sources | BBCSport | A549 | TCGA |
|---|---|---|---|---|
| ResNMTF (BiS/BiS) | 0.3209 | 0.2701 | 0.1914 | 0.3201 |
| ResNMTF (BiS/F) | **0.5143** | 0.5399 | 0.4594 | 0.5546 |
| ResNMTF (F/F) | 0.5022 | **0.6512** | **0.4632** | 0.5544 |
| NMTF | 0.4908 | 0.5931 | 0.3641 | **0.5833** |
| GFA | 0.4373 | 0.3474 | 0.3760 | 0.5475 |
| iSSVD | 0.2773 | 0.5154 | 0.1419 | 0.2039 |

### 5.2. Hyperparameter tuning

This section discusses the hyperparameter values and tuning implemented in the application of ResNMTF to both the synthetic and real data. Two alternative approaches for the two sets of data were implemented.

For all experiments concerning synthetic data, the biclusters are constructed to share common row clusters across rows and so restrictions between views are implemented via non-zero $\Phi$ with $\Psi = \Xi = 0$. Each pairwise restriction has the same weighting i.e. all $\phi_{vw} = \phi$ for some constant $\phi$. By varying $\phi$ across different scenarios, a suitable value of $\phi = 200$ for the synthetic data is found (see Section S2.4 for details). Stability analysis is performed with $\omega = 0.4$ (see Section 4.8 for the description of this hyperparameter). The value of $\omega$ applied on the synthetic datasets was selected based off preliminary investigations, however tuning of the stability analysis hyperparameter is investigated further on the real datasets.

In all the real datasets the samples are the shared dimension. The decision whether or not to transpose the views does affect results - for example, in the selection of the number of biclusters the bisilhouette score calculates the silhouette score of row clusters over columns corresponding to a bicluster. For 3Sources and BBCSport, matching terms to documents improves performance and so both datasets are transposed with restrictions enforced via non-zero $\Psi$. Similarly, for TCGA restriction is enforced between columns (individuals) via non-zero $\Psi$. In contrast, for the single-cell A549 dataset, performance is not improved by transposing and restriction is enforced via non-zero $\Phi$ for A549. As with the analysis of the synthetic data, restrictions are assumed to be equal between views with $\psi_{i,j} = \psi$ ($\phi_{i,j} = \phi$) for non-zero $\psi_{i,j}$ ($\phi_{i,j}$).

In the analysis of the real datasets, three strategies were considered for the tuning of the restriction and stability analysis hyperparameters. Restriction hyperparameters are optimised first, followed by the stability threshold. The first strategy is suitable for an unsupervised setting and uses the bisilhouette score to select the optimal hyperparameters - this is denoted ResNMTF (BiS/BiS). Secondly, the restriction hyperparameter is optimised via the bisilhouette score but $\omega$ is optimised using the F-score and is denoted by ResNMTF (BiS/F). Through this strategy, the ability of the bisilhouette score to correctly determine the optimal hyperparameters in the stability analysis and the restriction case, is disentangled. The final strategy considers the supervised setting where true labels are available. Here both hyperparameters are optimised via the F-score and the approach is denoted by ResNMTF (F /F). All implemented strategies do not require any user input, but ResNMTF (BiS/BiS) is the only strategy applicable in an unsupervised setting. As all meth-ods (including GFA and iSSVD) involve some degree of random initialisations the methods are applied on the real datasets with 5 different seeds. The F-score is used to select the best result between the repeats for NMTF, GFA and iSSVD, whereas for ResNMTF applications the measure used to optimise $\omega$ is used to select between the 5 repetitions (*i.e.* the F-score for ResNMTF (BiS/F)). As for most real datasets, only information about clusters present, and not the features that drive those clusters, is known. Relevance and the F-score for the real datasets are therefore calculated based on the row for A549, or column for 3sources, BBCSport and TCGA, clusters alone.

[1] http://erdos.ucd.ie/datasets/3sources.html (accessed 20th January 2024)
[2] https://github.com/kunzhan/SDSNE/blob/main/data/bbcsport_2view.mat (accessed 31st January 2024)
[3] https://github.com/sqjin/scAI/blob/master/data/data_A549.rda (accessed 28th January 2024)
[4] https://acgt.cs.tau.ac.il/multi_omic_benchmark/download.html (accessed 12th June 2025)

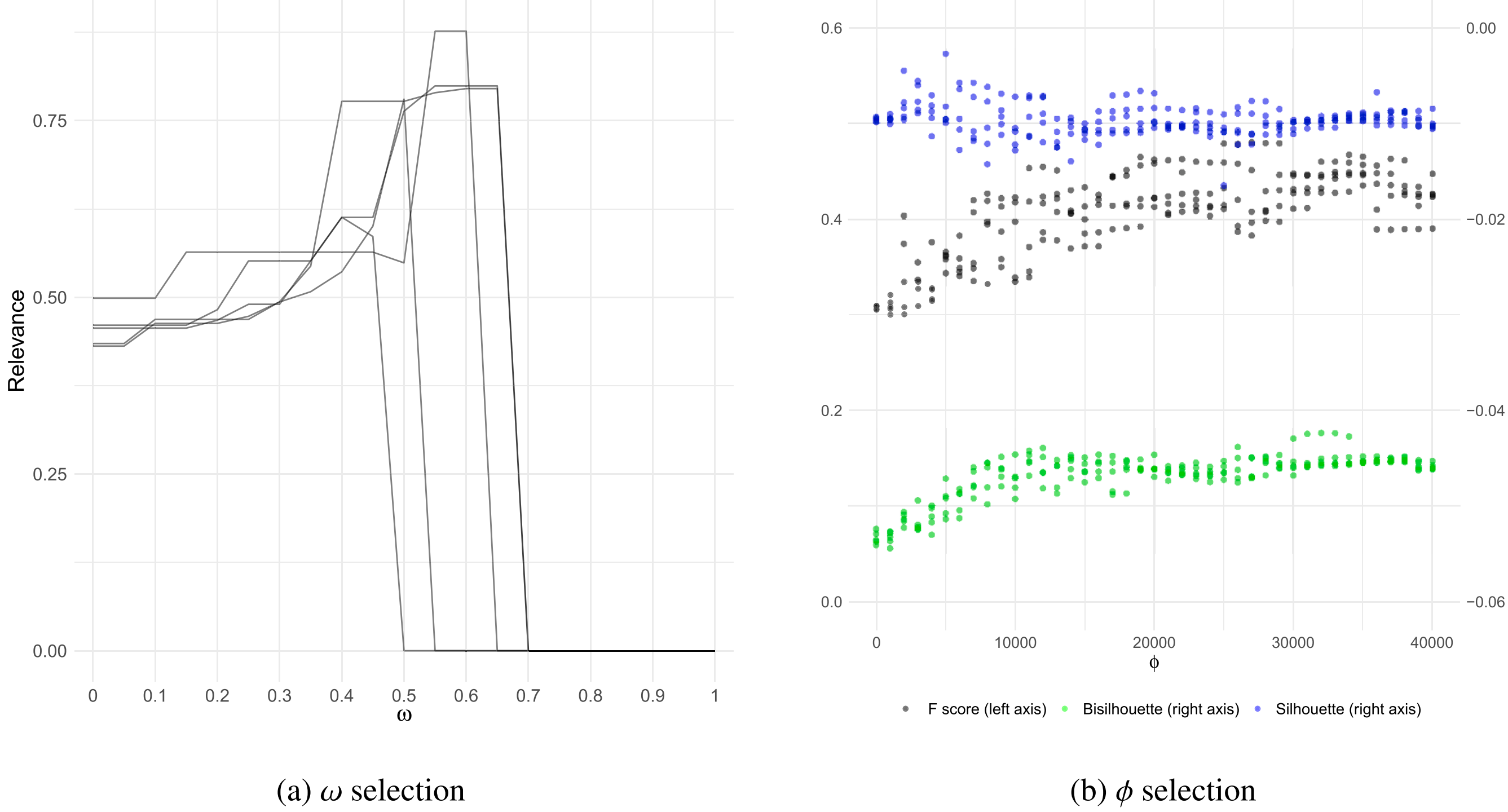

(a) $\omega$ selection

(b) $\phi$ selection

**Fig. 5.** Hyperparameter tuning: (a) relevance against $\omega$ on 3Sources dataset (using $\psi$ optimised by the bisilhouette score) and (b) F-score (left y axis, black), the bisilhouette score (right y axis, green) and the silhouette score (right y axis, blue) against $\phi$ on A549 dataset. In (a), each line corresponds to the results from one of 5 random seeds. In (b) the method was applied with each value of $\phi$ for 5 different seeds and all datapoints are shown. (For interpretation of the references to colour in this figure legend, the reader is referred to the web version of this article.)

### 5.3. Results

#### 5.3.1. ResNMTF performance

ResNMTF and NMTF are found to have similar performance and/or similar trends in their performance, with the former outperforming the latter in most cases (Fig. 4). Both approaches consistently outperfom both GFA and iSSVD. The performance of (Res)NMTF drops as the number of biclusters, and as expected, the level of noise, increases. Surprisingly, the performance of the methods is not found to improve as the number of individuals increases. NMTF outperforms ResNMTF when the number of views are increased and this is also seen when we consider 4 instead of 5 biclusters (Figure S10). One possible explanation for this is that the value of the restriction hyperparameter depends strongly on the number of views.

GFA's performance is largely invariant to all factors investigated apart from noise whilst iSSVD is mostly unable to detect biclusters when applied with the default parameter values. This is despite extensive effort to tune hyperparameters (Figures S5 and S6). Only once the signal of the synthetic data ($\mu$) is increased is iSSVD able to detect biclusters (Figure S7).

The presence (and indeed the rate) of overlapping biclusters has no discernable effect on the performance of ResNMTF, NMTF or GFA (Figure S12). Whilst the performance of ResNMTF/NMTF does fall as the rate of non-exhaustivity increases, both methods remain superior to GFA and iSSVD. The performance of iSSVD in contrast is found to be improved in the presence of overlapping and/or non-exhaustive biclusters.

For all datasets except TCGA, ResNMTF with the stability parameter optimised via F-score outperforms all other methods (Table 2). For TCGA, NMTF outperforms ResNMTF suggesting that looking at the views independently may better suit this dataset. In all other cases, integrating the multiple views through ResNMTF improves the biclustering performance. The bisilhouette scores for the results show ResNMTF to be outperforming all other methods in three out of the four datasets. For A549, it is the second-best performing. Naturally, ResNMTF with both hyperparameters optimised via the bisilhouette score gives the best result out of the ResNMTF approaches (Table S2).

For 3Sources, first tuning the restriction hyperparameter with the bisilhouette score improves on tuning with the F-score. However, the bisilhouette score appears unable to select the stability analysis parameter. The bisilhouette score may favour the removal of all but the strongest biclusters, returning only the most sufficiently compact and well separated. Fig. 5a illustrates that stability analysis performs as desired on the 3Sources dataset with the relevance of remaining biclusters increasing as the threshold increases.

#### 5.3.2. Bisilhouette score performance

Visually, the bisilhouette score demonstrates very similar trends to that of the F-score. This is confirmed quantitatively as the bisilhouette score correctly identifies the best performing method in 100 % of the scenarios considered in Fig. 4, whilst the ranking of methods by F-score and the bisilhouette score have a Pearson correlation of 0.944 (3sf) with disagreements only appearing between the two worst performing methods. This trend is further demonstrated in the remaining results presented in Section S2, most notably whilst tuning hyperparameters on iSSVD (Figures S5 and S6), in the presence of overlapping and non-exhaustive biclusters (Figure S12) and when considering data generated via an alternative process (Figure S13). This latter case demonstrates the use of the bisilhouette score outside of ResNMTF as a tool for tuning hyperparameters and comparing results in an unsupervised setting.

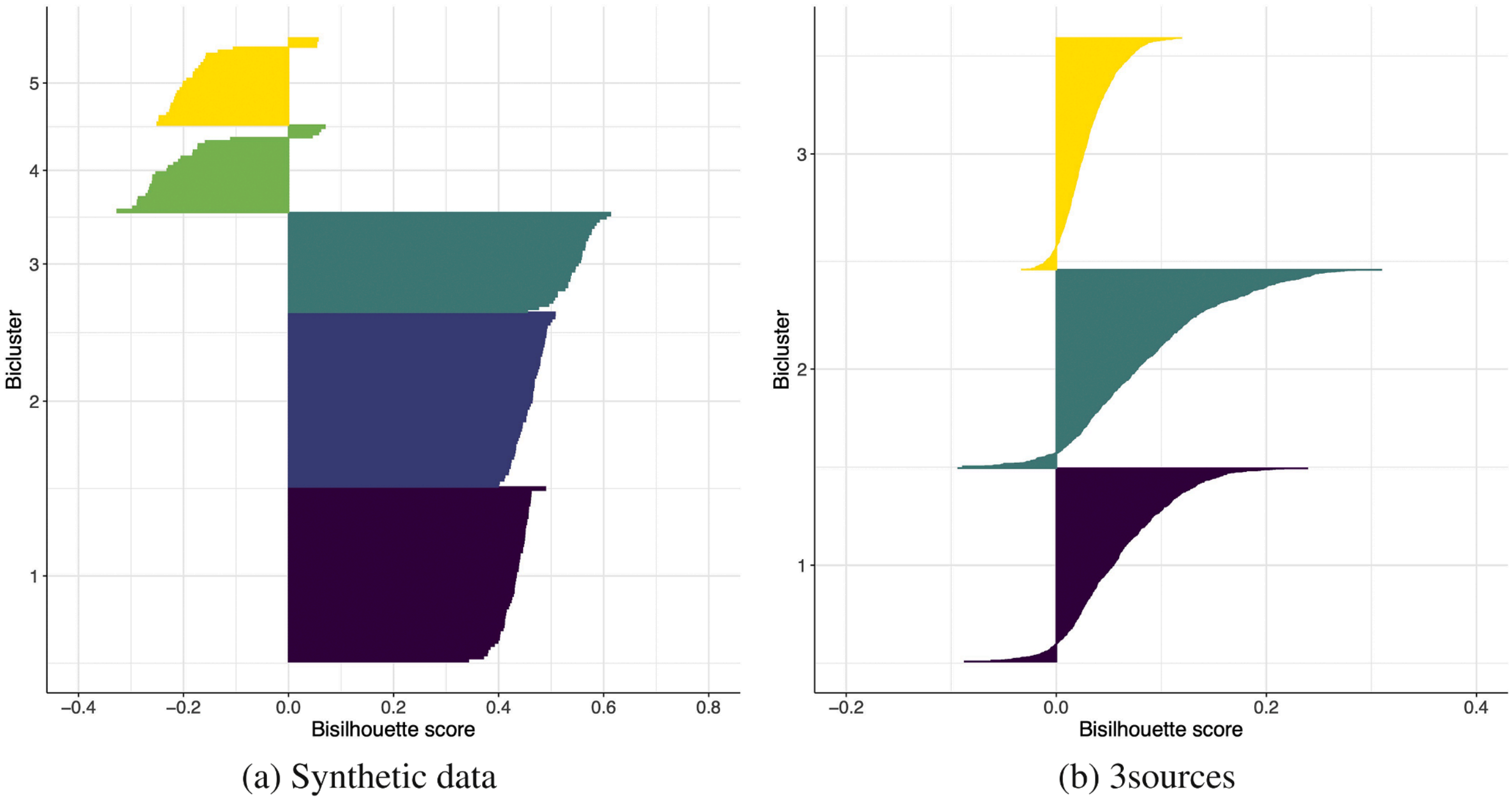

**Fig. 6.** Bisilhouette plots for view one of (a) synthetic data with 3 views and 5 biclusters using the true column clusters and 3 out of 5 of the correct row clusters with the remaining row clusters having been reassigned randomly, and (b) the 3sources dataset. The individual bisilhouette coefficients for each of the rows is shown along the x axis, with rows grouped into the assigned biclusters (denoted by different colours).

The analysis of the real datasets illustrates the suitability of the bisilhouette score for the selection of the restriction hyperparameters as previously discussed (Table 2). In particular, we see the trend shown by the bisilhouette score across $\phi$ values for the A549 dataset to be in agreement with the F-score. The two scores have a Pearson correlation equal to 0.722 (3sf; Fig. 5b). Notably the bisilhouette score is more closely aligned with the F-score compared to the traditional silhouette score. If the silhouette score is used instead to tune the restriction hyper parameter $\phi$ for dataset A549, the correlation between this intrinsic measure and the F-score falls to -0.0807 (3sf).

Overall, the bisilhouette scores of the final results from all methods show similar trends to the F-scores (Table S2). In particular, the rankings via F-score and the bisilhouette score of ResNMTF (BiS/F), ResNMTF (F/F) and NMTF agree in all cases. The two scores differ strongly in their ranking of the poorly performing methods, in particular for ResNMTF (BiS/BiS). This appears due to the bisilhouette score favouring a biclustering containing only the most well-defined bicluster (e.g. those that remain after all but the strongest bicluster are removed) over a biclustering containing several well-defined biclusters (with a necessarily lower average score).

Similar to the original silhouette score, the bisilhouette score can be used as a visualisation tool to aid in assessing biclusters. Fig. 6 illustrates two possible examples, where in Fig. 6a potentially incorrect biclusters are flagged, and in Fig. 6b known biclusters are verified. Further, the bisilhouette plots may be used to elicit information regarding the biclusters. For example, Fig. 6b suggests that all biclusters have captured some signal with bicluster 3 (in yellow) showing weaker differentiation. This is in agreement with the F-score contributions of each bicluster. The F-score of bicluster 3 is 0.314 (3sf) compared to 0.868 (3sf) and 0.755 (3sf) for biclusters 1 and 2 respectively.

## 6. Conclusion and discussion

This paper proposes an intrinsic measure for evaluating biclustering solutions, the bisilhouette score. The proposed score is a promising solution for tuning hyperparameters and selecting between biclusterings in an unsupervised setting. The bisilhouette score appears less adapt at selecting the stability analysis parameter than the restriction hyperparameter, within ResNMTF. This may be due to the bisilhouette score favouring the removal of all biclusters except that with the highest individual score. This suggests its lack of suitability in tuning a hyperparameter responsible for the removal or addition of biclusters. However, the promising results warrant further investigation into the use of the bisilhouette in tuning hyperparameters. Similarly to the original silhouette score, it provides visualisations which can aid in the assessment of biclustering quality. These qualities make the bisilhouette score an important tool in the application of biclustering methods in an unsupervised setting. Verma et al. [43] have previously extended the silhouette score to the bicluster case, however whereas their work involves a reformulation of the problem, our extension of the silhouette score does not change the underlying calculations - only the data they are performed on. Future work will compare the bisilhouette score to the measure by Verma et al. [43], as well as investigating further the ability of the bisilhouette score to tune the hyperparameters of other biclustering methods.

A novel multi-view biclustering approach, ResNMTF, is also proposed. In agreement with the literature the integration of multiple views improves performance providing further evidence of the benefit of multi-view data. As well as generally outperforming the single-view method, unrestricted NMTF, ResNMTF largely outperforms the multi-view biclustering approaches GFA and iSSVD, on both synthetic data and real world datasets. ResNMTF successfully identifies both overlapping and non-exhaustive biclusters, without pre-existing knowledge of the number of biclusters present, and is able to incorporate any combination of restrictions between views. Future work will investigate the convergent properties of ResNMTF. Whilst preliminary empirical results demonstrate convergence and suggest a rate dependent on $K$ (Figure S1), theoretical proofs of convergence of NMF algorithms with multiplicative update steps have been discussed in the literature to be difficult [44].

## CRediT authorship contribution statement

**Ella S. C. Orme:** Writing – review & editing, Writing – original draft, Visualization, Software, Methodology, Investigation, Formal analysis, Data curation, Conceptualization; **Theodoulos Rodosthenous:** Software, Methodology, Conceptualization; **Marina Evangelou:** Writing – review & editing, Supervision, Methodology, Conceptualization.

## Data availability

The real datasets analysed can be downloaded from; http://erdos.ucd.ie/datasets/3sources.html (3Sources, accessed 20th January 2024), https://github.com/sqjin/scAI/blob/master/data/data_A549.rda (A549, accessed 28th January 2024), https://github.com/kunzhan/SDSNE/blob/main/data/bbcsport_2view.mat (BBCSport, accessed 31st January 2024) and https://acgt.cs.tau.ac.il/multi_omic_benchmark/download.html (TCGA, accessed 12th June 2025).

## Declaration of competing interests

The authors declare that they have no known competing financial interests or personal relationships that could have appeared to influence the work reported in this paper.

## Acknowledgement

This work is funded by a doctoral scholarship from The Engineering and Physical Sciences Research Council (EPSRC).

## Supplementary material

Supplementary material associated with this article can be found in the online version at 10.1016/j.patcog.2025.112454